\documentclass[11pt,a4paper]{article}
\usepackage[latin1]{inputenc}
\usepackage{amsmath}
\usepackage{amsfonts}
\usepackage{amssymb}
\usepackage{epsfig}
\usepackage[lmargin=1in,rmargin=1in,tmargin=1.25in,bmargin=1.25in]{geometry}
\author{Sumit Ghosh}

\begin{document}
\title{Relativistic fermion on a ring: \\ Energy spectrum and persistent current}
\author{Sumit Ghosh\thanks{Email: sumit.ghosh@bose.res.in}\\
S N Bose National Centre for Basic Sciences \\ Block JD, Sector III, Salt Lake, Kolkata 700098}
\date{}
\maketitle

\begin{abstract}
The energy and persistent current spectra for a relativistic fermion on a ring is studied in details. The nonlinear nature of persistent current in relativistic regime and its dependence on particle mass and ring radius are analysed thoroughly. For a particular ring radius we find the existence of a critical mass at which the single ring current doesn't depend on the flux. In lower mass regime the total current spectrum shows plateaus at different height which appears periodically. The susceptibility as well shows periodic nature with amplitude depending on particle mass. As we move from higher mass to lower mass regime, we find that the system turns into paramagnetic from diamagnetic. We also show that same behaviour is observed if one vary the radius of the ring for a fixed particle mass. Hence the larger ring will be diamagnetic while the smaller one will be paramagnetic. Finally we propose an experiment to verify our findings.  
\end{abstract}

\section{Introduction}

The existence of a persistent current in a normal metallic ring threaded by a magnetic flux has stimulated a large number of studies in last few decades \cite{Butt1}-\cite{Kulik}. The current is periodic in flux like the current in a superconducting ring \cite{Byers}-\cite{Bloch2}. Due to the recent advances in fabrication techniques, it has become possible to observe this current experimentally \cite{Levy}-\cite{Jayich} in mesoscopic systems. Such current is also found in carbon nanotube rings \cite{Szopa}-\cite{Chena}. In these carbon based structures electrons behaves as massless Dirac particles and they require a relativistic description. Although there have been a significant amount of work on the energy levels and persistent current of relativistic fermion regarding graphene \cite{Ino}-\cite{Huang}, topological insulator \cite{Mich} and for  neutrino billiards \cite{Berry} and Dirac electron \cite{Papp}, there is still lack of a complete description of persistent current for relativistic fermions and its variation with particle mass. Besides, it is well known that superconducting as well as mesoscopic rings are perfect diamagnets, but carbon nanotube tori \cite{Tam} are found to posses a positive susceptibility. There is no vivid discussion on how the transition occurs.   

In this paper we have studied the nature of energy and persistent current spectra for a relativistic particle on a one dimensional ring. We describe the scenario for all the mass regime and explain the massless case as well. For heavy fermions our findings are in good agreement with the nonrelativistic behaviour \cite{Gefen1}. Interesting features are revealed as we move towards the relativistic regime, i.e. for low mass. The flux dependence of the current becomes more and more nonlinear as we decrease the mass. For a particular ring radius we find the existence of a critical mass at which the current becomes constant in flux. The value of this current is same as the current produced by a massless particle and it appears to be the lower cutoff for current as we further decrease the mass. This nature is prominent for massless particle as well. We further study the nature of the total current and find that it shows periodic plateaus. It becomes a multiple step function at critical mass and becomes discrete below that mass. The same periodicity is also observed in the susceptibility. More interestingly, we find that its amplitude changes from negative to a positive value as we decrease the mass. This is further clarified by studying the behaviour of the susceptibility with respect to mass at a particular flux value. Same behaviour is observed if we keep the mass fixed and treat the ring radius as parameter.     

The organisation of the paper is as follows. In section 2 we give a brief description about the Dirac equation and its solution on a circle. Then we obtain the energy spectrum in presence of a Aharonov-Bohm flux. We derive the resulting persistent current in section 3 and show its behaviour for different mass range specially for a massless fermion. In section 4 we discuss the behaviour of total current and in section 5 the nature of susceptibility is discussed. Section 6 contains a proposed experiment to verify the result. The discussion is kept in section 7.  

\section{Relativistic particle on a one dimensional ring}

We start with a very brief description of Dirac equation on a circle. Then we will derive its solution in presence of an Aharonov-Bohm flux and obtain the energy spectrum.

Dirac equation in two dimensions is given by \cite{Thaller}
\begin{eqnarray}
(c\vec{\sigma}\cdot\vec{p} + \sigma_3 m c^2)\psi = E \psi
\end{eqnarray}
where $\psi = (\zeta_1,\zeta_2)^T$ is a two component spinor. $\vec{\sigma} =\sigma_1\hat{x} + \sigma_2\hat{y}$ where $\sigma_1$, $\sigma_2$ and $\sigma_3$ are Pauli matrices. Due to the symmetry of the problem, it would be easier to work with polar coordinate ($x=Rcos(\varphi), y=Rsin(\varphi)$: $R$ is the ring radius) in which the Dirac Hamiltonian becomes
\begin{eqnarray}
\vec{\sigma}\cdot\vec{p} + \sigma_3 mc^2 = \begin{pmatrix} mc^2 & -\frac{c\hbar}{R}e^{-i\varphi} \frac{\partial}{\partial \varphi} \\ \frac{c\hbar}{R}e^{i\varphi} \frac{\partial}{\partial \varphi} & -mc^2 \end{pmatrix} = c\sigma_\varphi p_\varphi + \sigma_3 mc^2
\label{HD}
\end{eqnarray}
where $\sigma_\varphi = \begin{pmatrix} 0 & -i\frac{1}{R}e^{-i\varphi} \\ i\frac{1}{R}e^{i\varphi} & 0 \end{pmatrix}$ and $p_\varphi = -i\hbar \frac{\partial}{\partial \varphi}$. Note that (\ref{HD}) is similar to the Hamiltonian used in \cite{Papp}. They have used $4\times4$ Dirac matrices whereas we use here $2\times2$ representation which finally leads to the same dispersion relation. It is worth mentioning that one should be careful while deriving a Dirac Hamiltonian for a lower dimension. Remember that number of Dirac matrices required for a $d+1$ (space-time) dimensional space is $d+1$ and they must satisfy the corresponding Dirac algebra. One can start from a two dimensional Dirac Hamiltonian and replace the radial variable dependent terms with their expectation values considering a vanishing width of the radial wave function (\cite{Meij},\cite{Peet}). In this way although it is possible to reduce a spatial dimension, but the resulting Hamiltonian may still need three Dirac matrices\footnote{One can check that by using  $\langle r \rangle=0$ and $\langle \frac{\partial}{\partial r} \rangle=-\frac{1}{2R}$ \cite{Meij} in a two dimensional polar Dirac equation.}  and that will be a contradiction about the dimensionality of the system. A better way would be to use the constraint relations and obtain the Hamiltonian in terms of the generalised coordinate which we have done here. The Hamiltonian (\ref{HD}) we derive here involves only two Dirac matrices ($\sigma_\varphi,\sigma_3$) which is in tone with the one dimensionality of  the system. 

\begin{figure}[h]
\centering
\epsfig{file=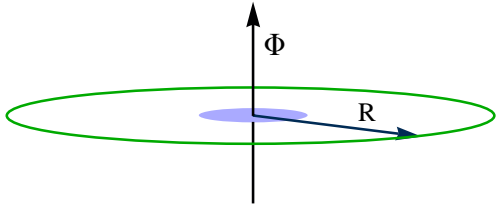,scale=0.5}
\caption{A one dimensional ring of radius R threaded by a magnetic flux $\Phi$. The field is finite only within the shaded region.}
\label{ring}
\end{figure}

Now let us consider a perpendicular magnetic field piercing the ring ($fig.$\ref{ring}). A suitable choice for vector potential (under symmetric gauge) is
\begin{eqnarray}
\vec{A} = \frac{\Phi}{2\pi R} \hat{\varphi}
\label{A}
\end{eqnarray}
where $\Phi$ is the total flux passing through the ring.

The energy of the system is given by \cite{Papp}
\begin{eqnarray}
E_l = \sqrt{m^2c^4+\left(\frac{c\hbar}{R}\right)^2 (l+\frac{\Phi}{\Phi_0})(l+\frac{\Phi}{\Phi_0}-1)}
\label{ep}
\end{eqnarray}
where $l=0,\pm1,\pm2,\cdots$ and $\Phi_0 = \frac{ch}{e}$ is the flux quanta.
\begin{figure}[h]
\centering
\epsfig{file=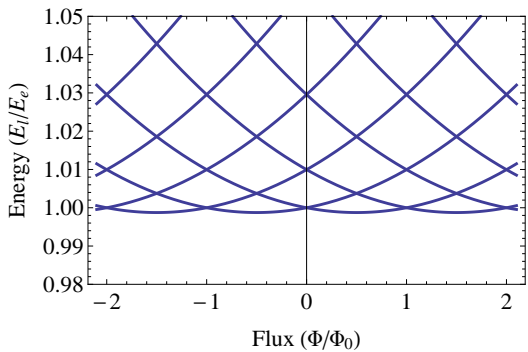,scale=.75}
\caption{Energy as a function of flux. Here we have used $\hbar=c=e=m_e=1$ ($e$ and $m_e$ being electronic charge and mass. The units of mass, length and energy are $m_e$, $r_e=\frac{\hbar}{m_e c}$ and $E_e=m_ec^2$). The ring radius and particle mass are $R=10r_e$ and $m=1m_e$ respectively. The energy minima are at $\Phi/\Phi_0=l+1/2$. This shift is a signature of relativistic correction.}
\label{energy}
\end{figure}

One can see ($fig.$\ref{energy}) that the energy varies parabolically with respect to $\Phi$ and has a periodicity of $\Phi_0$. This is similar to the case of a nonrelativistic particle \cite{Gefen1}. The only difference is the fact that the minima of the energy is at half integral multiple of $\Phi_0$. This shift by $\Phi_0/2$ appears due to relativistic correction \cite{Papp}.

\section{Persistent current for a Dirac particle}

We will now discuss about the nature of energy spectrum and resulting persistent current for different mass regime. 

Persistent current is the perpetual current flowing in a superconducting or normal metallic ring in presence of a Aharonov-Bohm flux. It is given by the flux derivative of the energy \cite{Gefen1,Bloch2}
\begin{eqnarray}
I_l = -c \frac{\partial E_l}{\partial \Phi}
\label{id}
\end{eqnarray}
For a Dirac particle on a one dimensional ring whose energy is given by (\ref{ep}), the persistent current is  
\begin{eqnarray}
I_l = -\frac{e\hbar}{2\pi mR^2} \frac{l+\frac{\Phi}{\Phi_0}-\frac{1}{2}}{\sqrt{1+\left(\frac{\hbar}{mcR}\right)^2 \left(l+\frac{\Phi}{\Phi_0}\right)\left(l+\frac{\Phi}{\Phi_0}-1\right)}}
\label{il}
\end{eqnarray}

\begin{figure}[h]
\centering
\epsfig{file=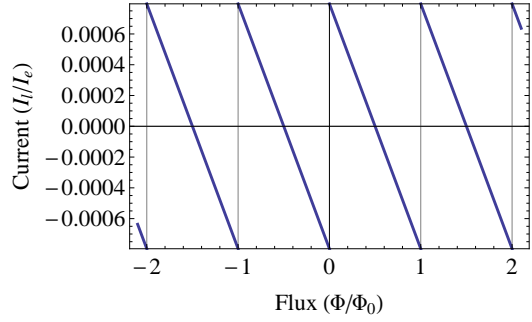,scale=.75}
\caption{Variation of current with flux for $R=10r_e$ and $m=1m_e$. Here we have used the same unit system as $fig.$\ref{energy} and current is measured in units of $I_e=\frac{e c}{r_e}$.}
\label{current}
\end{figure}

One can regain the nonrelativistic result for $m \rightarrow \infty$. At this limit the current is given by
\begin{eqnarray}
I_l\vert_{m\rightarrow \infty} = -\frac{e\hbar}{2\pi mR^2} \left(l+\frac{\Phi}{\Phi_0}-\frac{1}{2}\right)
\label{inr}
\end{eqnarray}

The expression looks similar to the nonrelativistic current \cite{Gefen1}. The only difference is that like the energy spectrum, it also undergoes a shift by $\Phi_0/2$ with respect to the flux. However the behaviour of current at relativistic regime ($m\rightarrow 0$) is rather complicated. To have an idea about the flux dependence of the current we will study the single particle susceptibility $\Omega_l(\kappa,\Phi)$ defined as
\begin{eqnarray}
\Omega_l(\Phi,\kappa) = \frac{\partial I_l}{\partial \Phi}  = \frac{I_0}{\Phi_0} \frac{\kappa -4}{4\left[ 1+\kappa \left(l+\frac{\Phi}{\Phi_0}\right)\left(l+\frac{\Phi}{\Phi_0}-1\right) \right]^{3/2}}
\label{slope}
\end{eqnarray}
where 
\begin{eqnarray}
I_0 &=& \frac{e\hbar}{2\pi mR^2} \label{i0} \\
\kappa &=& \left(\frac{\hbar}{mcR}\right)^2 \label{kappa}
\end{eqnarray}

\begin{figure}[h]
\centering
\epsfig{file=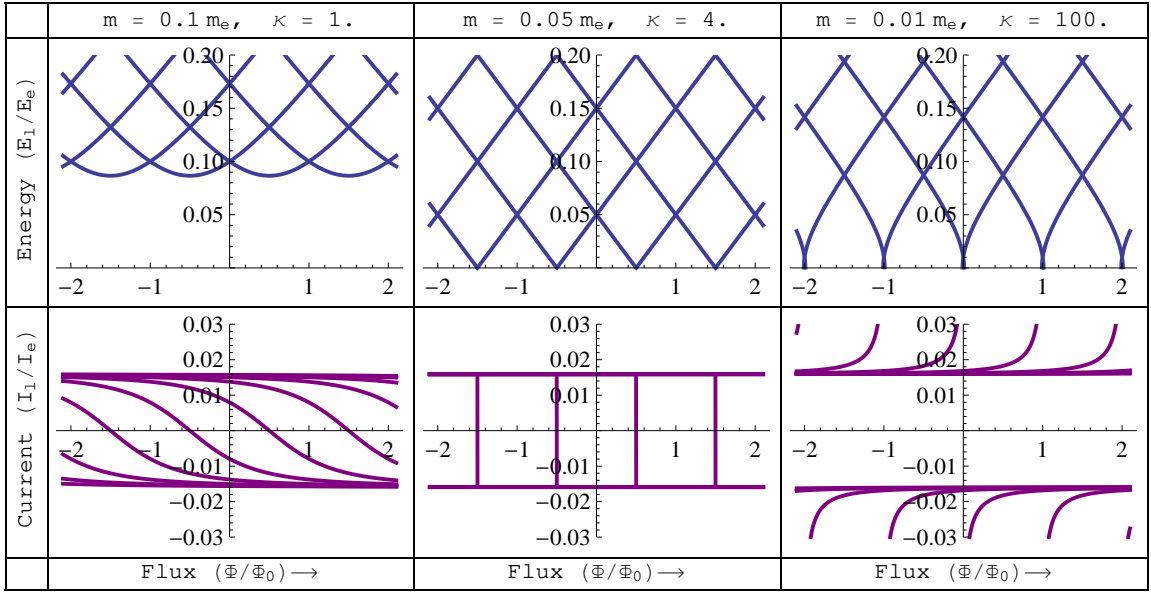,scale=1}
\caption{Nature of the energy and current for different values of $\kappa$. Here we use ($\hbar=c=e=m_e=1$) $R=10r_e$ and vary the mass. We have taken $m=0.1m_e~(\kappa=1)$, $m=0.05m_e~(\kappa=4)$ and $m=0.01m_e~(\kappa=100)$. $m=0.05m_e$ is the critical mass (\ref{mcr}) at which energy becomes linear in flux and current attains a constant value $I_{sat}$ (\ref{isat}). For $\kappa>4$, $I_{sat}$ is the lower cutoff of the current.}
\label{table}
\end{figure}

Now we will discuss the behaviour of $\Omega_l(\Phi,\kappa)$ for different range of $\Phi$ and $\kappa$.
\begin{itemize}
\item[\it a.] $\underline{\kappa \rightarrow 0 :}$ $\Omega_l(\Phi,\kappa)\rightarrow -I_0/\Phi_0$ which is negative, constant and independent of $l$. Hence the current will vary linearly with flux for all $l$ values. This is actually the nonrelativistic ($m\rightarrow \infty$) limit.
\item[\it b.]$\underline{\kappa=4 :}$ $\Omega_l(\Phi,\kappa)=0$, i.e. the current will not depend on the flux. We call the corresponding mass to be the critical mass for radius $R$, given by
\begin{equation}
m^R_{cri} = \frac{\hbar}{2cR}
\label{mcr}
\end{equation}

The constant current is given by 
\begin{eqnarray}
I_{sat}= -\frac{I_0}{\kappa} = -\frac{ec}{2\pi R}
\label{isat}
\end{eqnarray}
This is actually current generated by a massless particle with charge $e$.
\item[\it c.]$\underline{\kappa>4 :}$ This is the low mass limit. At $\Phi\rightarrow \infty$ limit the current becomes constant, given by (\ref{isat}). For $\kappa>4$, $\Omega_l(\Phi,\kappa)$ blows up at 
\begin{eqnarray}
\left. \frac{\Phi}{\Phi_0}\right|_{div} = \left( \frac{1}{2} -l\right) \pm \frac{1}{2} \sqrt{1-\frac{4}{\kappa}} 
\label{pdiv}
\end{eqnarray}
which are actually the roots of the denominator of (\ref{slope}). The divergence in $\Omega_l(\Phi,\kappa)$ suggests a discontinuity in current. For $\kappa \rightarrow \infty$ ($m=0$, i.e. massless limit) the divergence occurs at $\frac{\Phi}{\Phi_0} =-l$ and $\frac{\Phi}{\Phi_0} =-l+1$. Note that within these two diverging limits the energy becomes a imaginary quantity and hence the corresponding states will be decaying states. Appearance of such non hermiticity of Dirac Hamiltonian is well known in curved space \cite{Xing}. In our case it is an artefact of the geometry of the system.
\end{itemize}   
The nature of energy and current in these three $\kappa$ region is shown in $fig.$\ref{table}. $Fig$ \ref{energy} and \ref{current} show the energy and current spectra for $\kappa=0.01$ ($R=10r_e,m=1m_e$) which is an example for $\kappa\rightarrow 0$ regime.

One can also observe the same nature of current by keeping $m$ fixed and varying $R$ as a parameter. In that case there would be a critical radius given by
\begin{equation}
R^m_{cri} = \frac{\hbar}{2mc}
\label{rcr}
\end{equation}
at which the current becomes constant. This is actually half of the reduced Compton wavelength of the particle. However by tuning the radius it is not possible study the phenomena for massless particles.

\subsection{Persistent current for a massless Dirac particle}

As a special case, we will discuss here the nature of persistent current caused by a massless particle. The energy of such particle can be obtained by putting $m=0$ in (\ref{ep}), and is given by
\begin{eqnarray}
E^0_l = \left(\frac{c\hbar}{R}\right) \sqrt{(l+\frac{\Phi}{\Phi_0})(l+\frac{\Phi}{\Phi_0}-1)}
\label{ep0}
\end{eqnarray}

\begin{figure}[!h]
\centering
\epsfig{file=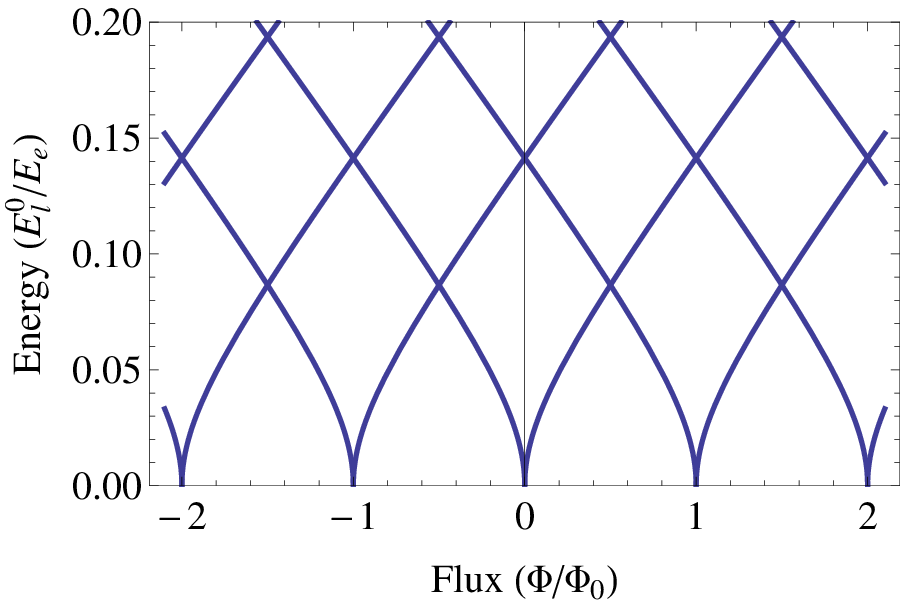,scale=.75}
\epsfig{file=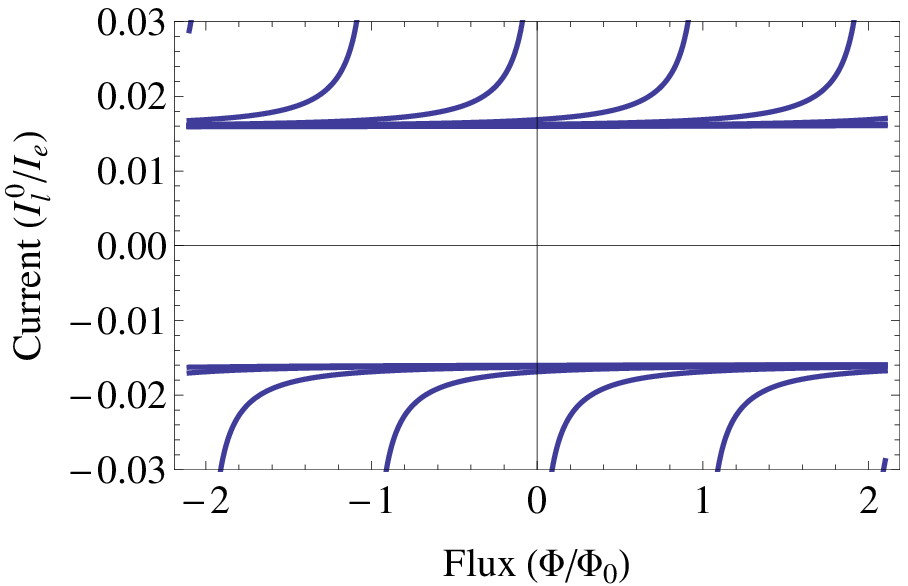,scale=.75}
\caption{Nature of energy and current for massless particles. The lower cutoff value for current is given by $ec/2\pi R$. Here we used R=10$r_e$ with the same units as $fig.$\ref{energy} and $fig.$\ref{current}.}
\label{mass0}
\end{figure}

Corresponding persistent current by definition (\ref{id}) is 
\begin{eqnarray}
I^0_l = -\frac{ec}{2\pi R} \frac{l+\frac{\Phi}{\Phi_0}-\frac{1}{2}}{\sqrt{ \left(l+\frac{\Phi}{\Phi_0}\right)\left(l+\frac{\Phi}{\Phi_0}-1\right)}}
\label{il0}
\end{eqnarray} 
The variation of $E^0_l$ and $I^0_l$ with flux is shown in $fig.$\ref{mass0}.

From the denominator of (\ref{il0}) one can readily see that for $\Phi/\Phi_0=-l$ and $\Phi/\Phi_0=-l+1$ the current diverges. This is something that does not happen for a nonrelativistic particle. At asymptotic region ($\Phi/\Phi_0 \rightarrow \infty$) the current becomes constant ($fig.$\ref{mass0}) and is given by (\ref{isat}) which is the current produced by a massless particle with charge $e$ circling in a ring of radius $R$.

\section{Total current for odd and even number of particles}

\begin{figure}[!h]
\centering
\epsfig{file=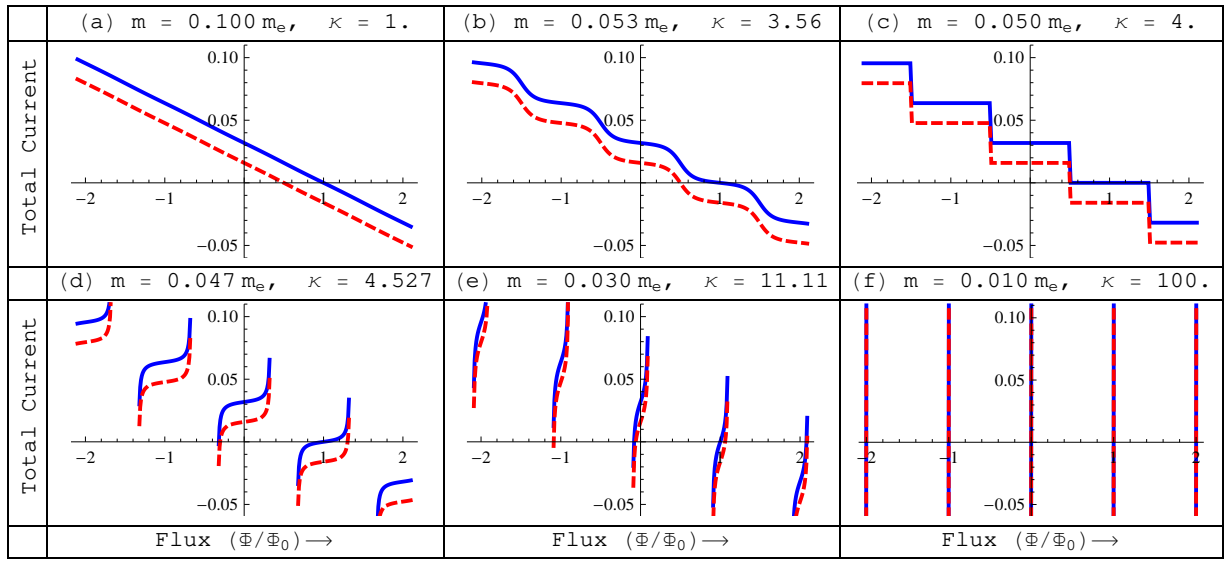,scale=1}
\caption{Total current (measured in units of $I_e$ ($fig.$\ref{current})) for 100 (Blue) and 101(Red dashed) particles in a ring with R=10$r_e$ ($\hbar=c=e=m_e=1$). Masses and corresponding $\kappa$ values are written on the top of each plot. For large mass ($a$) total current depends linearly on flux. As we decrease the mass ($b$) the dependence becomes nonlinear. For critical mass (\ref{mcr}), total current is a step function of flux ($c$), the step heights being multiple of $I_{sat}$ (\ref{isat}). For lower mass ($d$), total current tends to diverge at values given by (\ref{pdiv}), the mean position of the plateaus being multiple of $I_{sat}$. On further decrease of mass ($e$) the spectrum becomes narrower and the divergences come closer to the integral flux values. As we go to the zero mass limit ($f$), total current appears as sharp pulses at integral values of flux.}
\label{itot}
\end{figure}

From (\ref{il0}) one can understand that an exact analytic expression for total current is not possible for our case\footnote{For an approximate result at large mass limit see \cite{Papp}.}. We consider 100 particles in a ring with radius $R=10r_e$ and numerically evaluate the total current for different $m$ values. While doing so we consider only Pauli's exclusion principle and neglected any other kind of interaction among the particles. The result for both even and odd number of particles is shown in $fig.$\ref{itot}.

From $fig.$\ref{itot} we see that for large mass, total current varies linearly with flux ($fig.$\ref{itot}$a$). This is consistent with the nonrelativistic result. As we decrease the mass, relativistic effects starts dominating and the current starts showing nonlinear behaviour ($fig.$\ref{itot}$b$). This is manifested by the periodic appearance of flat regions in $fig.$\ref{itot}$b$. At $m=m^R_{cri}$ (\ref{mcr}) ($\kappa=4$) total current becomes a multiple step function of flux ($fig.$\ref{itot}$c$), the step heights being integral multiple of $I_{sat}$ (\ref{isat}) and each step being $2I_{sat}$ high. As we further decrease the mass ($\kappa>4$) the total current starts showing divergences ($fig.$\ref{itot}$d$). The point of divergence is given by (\ref{pdiv}). The average of a plateau between two divergences is at some multiple of $I_{sat}$ and located at integral values of flux. The width of the plateau decreases with the decrease of mass ($fig.$\ref{itot}$e$). As we go to the massless limit ($fig.$\ref{itot}$f$), the total current appears as sharp pulses at integral values of flux.   

\section{Susceptibility of a many electron ring}
The total susceptibility is evaluated by taking the flux derivative of the total current. It is well known that superconducting rings are perfect diamagnets and so are the mesoscopic metalic rings. Let see the nature of the susceptibility for the relativistic fermions.

\begin{figure}[h!]
\centering
\epsfig{file=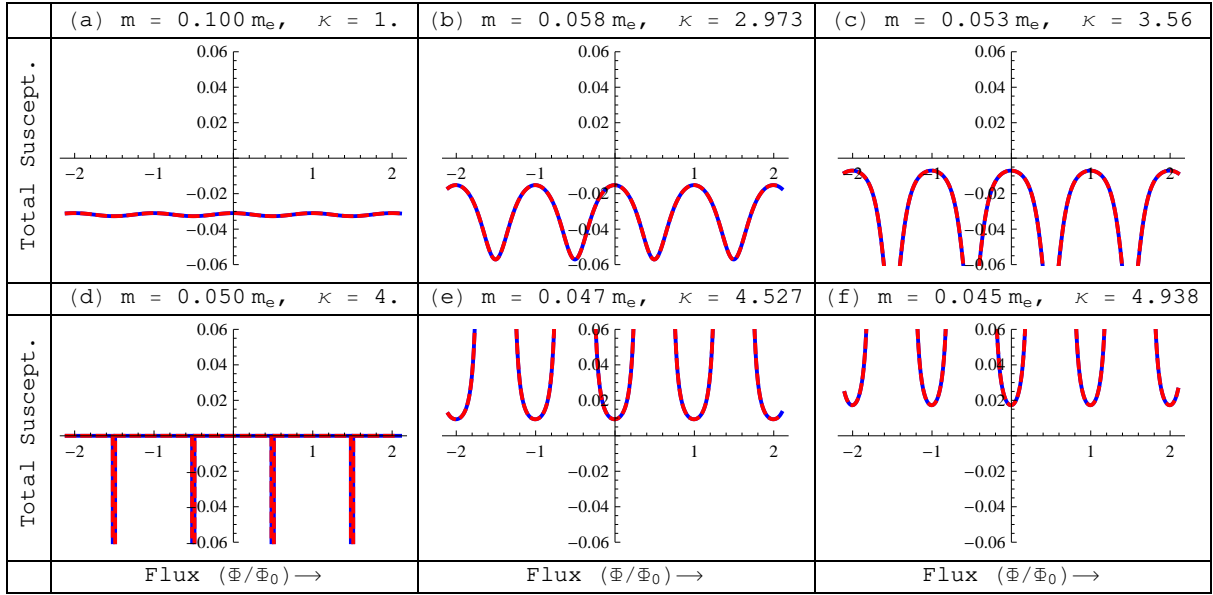,scale=1}
\caption{Nature of susceptibility for a ring with radius R=10$r_e$ at different masses which are written on the top of every plot. We have used the same units as $fig.$\ref{energy} in which the unit of susceptibility is $\Omega_e=\frac{e I_e}{r_e E_e}$. For large mass susceptibility is negative ($a$) and almost constant. It shows periodic oscillation with a negative amplitude ($b,c$) for $m>m_{cri}$ and becomes zero at critical mass ($d$). For lower mass it again shows oscillation ($e,f$) but  with positive amplitude. The blue and red dashed lines correspond to even and odd number of electrons which overlaps with each other.}
\label{susc}
\end{figure} 

The change of the magnetic susceptibility with mass is quite interesting for relativistic fermions ($fig.$\ref{susc}). For large mass the susceptibility is negative and almost constant ($fig.$\ref{susc}$a$). The ring behaves as a diamagnet like a superconducting ring. As we decrease mass it starts showing periodic oscillation with negative amplitude ($fig.$\ref{susc}$(b,c)$) with period $\Phi_0$. The maxima are located at integral flux values. The susceptibility becomes zero at critical mass $fig.$\ref{susc}$d$ with a sharp divergence at half integral values of flux which are due to change of step height in flux-current relationship curve $fig.$\ref{itot}$c$. As we further decrease the mass, the periodic oscillation appears again, but this time with positive amplitude $fig.$\ref{susc}$(e,f)$. The susceptibility has a minimum value at integral flux and encounters a positive divergence at half integral values of flux. The positive magnetic susceptibility is also reported for carbon nanotube torus \cite{Tam} where the electrons behave as massless Dirac fermions. To understand the the dependence of susceptibility on particle mass it is better to focus at the vicinity of an integral value of flux. Here we choose $\Phi=0$ and study the variation of susceptibility both with respect to mass and ring radius.
\begin{figure}[!h]
\centering
\epsfig{file=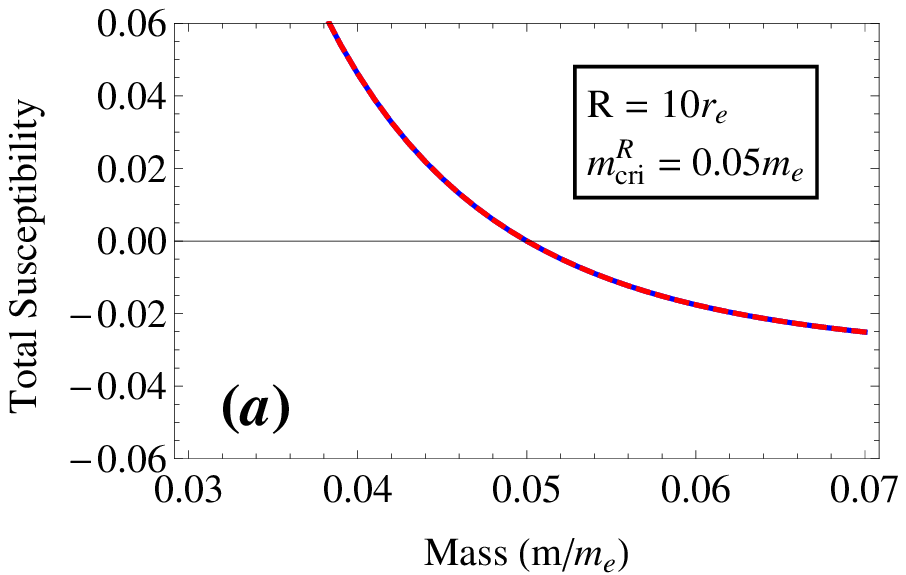,scale=.8}
\epsfig{file=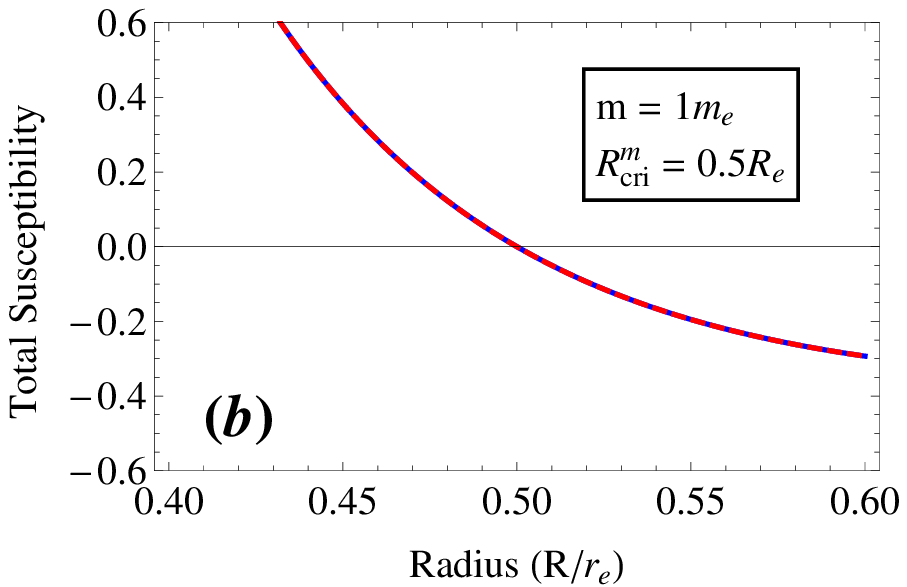,scale=.8}
\caption{Variation of susceptibility at $\frac{\Phi}{\Phi_0}=0$ with respect to \textbf{(\textit{a})} mass ($m$) for a ring with $R=10r_e$ and \textbf{(\textit{b})} ring radius ($R$) for $m=1m_e$. We have used the same units as $fig.$\ref{energy} and susceptibility is measured in units of $\Omega_e$ ($fig.$\ref{susc}). In both cases the susceptibility curve cut the x axis at corresponding critical value, i.e. at $m=m^R_{cri}=0.05m_e$ in $(a)$ and at $R=R^m_{cri}=0.5R_e$ in $(b)$. Both these points corresponds $\kappa=4$.}
\label{susm}
\end{figure}   

One can have a better idea about the change of susceptibility with respect to mass from $fig.$\ref{susm}$a$. For large mass we see that the susceptibility is negative and varies little with mass. This is the nonrelativistic limit where the rings behave as diamagnets. The susceptibility increases with the decrease of mass and becomes zero at critical mass. On further decrease of mass the susceptibility becomes positive and keeps increasing as we proceed to the zero mass limit.  The many particle fermionic ring thus makes a smooth diamagnetic to paramagnetic transition with decrease of mass. 
One must have noticed that for single particle susceptibility the real governing parameter is $\kappa$ (\ref{slope}) which is proportional to the inverse of the product of $m$ and $R$ (\ref{kappa}). Hence same behaviour of total susceptibility can be observed if one keeps $m$ fixed and tune $R$. We choose a fixed value of the particle mass ($m=1m_e$) and plot the total susceptibility against the ring radius ($R$) ($fig.$\ref{susm}$b$) and we find the nature to be similar to that of the total susceptibility - mass curve ($fig.$\ref{susm}$a$). The zero susceptibility occurs at $R=R^m_{cri}$ which is half the reduced Compton wavelength of the particle (\ref{rcr}). Thus from $fig.$\ref{susm}$b$ we can easily see that for larger radius the ring will be diamagnetic where as smaller rings will be paramagnetic.
 

\section{Possible experiment}

In this paper we have discussed everything keeping mass as a parameter. But for experimental purpose variation of particle mass is not a very useful idea. One can alternatively fix the mass and vary the radius ($R$) of the ring. The critical radius ($R^m_{cri}$) (\ref{rcr}) at which the single particle current becomes constant ($I_{sat}$) and consequently susceptibility becomes zero is found to be half of the reduced Compton wavelength of the particle. But designing such small ring is a great challenge and in this way we can't go to the massless limit either. On the other hand doped graphene can be a good testing ground for such experiment. On a pure graphene sheet electrons behave as massless Dirac fermions. In case of a doped graphene there is band gap \cite{Ohta}-\cite{Papagno}, which corresponds a massive Dirac fermion. The band-gap can be controlled by the doping concentration. Thus instead of mass one can use doping concentration as the driving parameter. Another way to introduce a mass gap in energy spectrum is to grow the graphene sheet on different substrate \cite{Zhou}. By controlling the mass gap it may be possible to study the nature of the persistent current. However confining electron in a one dimensional ring in graphene would be a great challenge. Due to Klein tunnelling, potential barriers will not be sufficient for confinement. The edge of a graphene disk can provide a one dimensional path, but due to edge effect the physics would be different from that of a simple Dirac fermion.   

\section{Conclusion}
In this paper we thoroughly study the nature of persistent current for  relativistic fermion. From our analysis it is clear that the current is a sensitive function of the particle mass as well as the radius of the ring. For large mass the total current is linear in flux with a constant negative susceptibility. But as we move to low mass regime the behaviour is more and more nonlinear which is clearly shown in $fig.$\ref{itot}. We find the existence of a critical mass ($m^R_{cri}$) below which the energy becomes imaginary which indicates towards unstable states for certain range of flux. The current shows a divergence within this range. Even more interesting is the nature of the susceptibility. For large mass it is negative and constant like a nonrelativistic case. As we go to low mass regime, the susceptibility appears to be a periodic function of flux with periodicity $\Phi_0$ ($fig.$\ref{susc}). Not only that, its amplitude also increases with the decrease of mass. At $m=m^R_{cri}$ ($\kappa=4$) the susceptibility becomes zero ($fig.$\ref{susc}$d$) and gradually turns positive with further decrease in mass. The same behaviour can also be observed if we fix the mass and vary the ring radius (\ref{susm}$b$). We show that a ring with bigger radius ($R>R^m_{cri}$, where $R^m_{cri}$ is half the reduced Compton wavelength of the particle ($fig.$\ref{rcr})) is diamagnetic while that with smaller one ($R<R^m_{cri}$) is paramagnetic.  

\section{Acknowledgement}
The author likes to thank Alexander Altland for helpful comments. The work is financially supported by Council of Scientific and Industrial Research, India.


\end{document}